# Sequence Classification of the Limit Order Book using Recurrent Neural Networks

Matthew Dixon[1]

[1]Stuart School of Business,
Illinois Institute of Technology,
10 West 35th Street,
Chicago, IL 60616,
matthew.dixon@stuart.iit.edu

July 14, 2017

**Abstract**

Recurrent neural networks (RNNs) are types of artificial neural networks (ANNs) that are well suited to forecasting and sequence classification. They have been applied extensively to forecasting univariate financial time series, however their application to high frequency trading has not been previously considered. This paper solves a sequence classification problem in which a short sequence of observations of limit order book depths and market orders is used to predict a next event price-flip. The capability to adjust quotes according to this prediction reduces the likelihood of adverse price selection. Our results demonstrate the ability of the RNN to capture the non-linear relationship between the near-term price-flips and a spatio-temporal representation of the limit order book. The RNN compares favorably with other classifiers, including a linear Kalman filter, using S&P500 E-mini futures level II data over the month of August 2016. Further results assess the effect of retraining the RNN daily and the sensitivity of the performance to trade latency.

## 1 Introduction

Many of the challenges facing methods of financial econometrics include non-stationarity, non-linearity or noisiness of the time series. The application of artificial neural networks (ANNs) to uni-variate financial time series methods are well documented (Faraway and Chatfield, 1998; Refenes, 1994; Trippi and DeSieno, 1992; Kaastra and Boyd, 1995). Their proneness to over-fitting, convergence problems, and difficulty of implementation raised concerns that hampered their early adoption by finance industry practitioners in predicting low frequency volatility or price changes.

Modeling and analysis of time-varying signals is an important subfield of signal processing. One of the most successful algorithms to arise for dynamic systems is the classical Kalman filter (KF). KFs belong to a family of Bayes' filters, which include particle filtering and Hidden Markov Models. The basic approach performs minimum mean square error estimation of the hidden state of a time-varying linear system. There is ample evidence to suggest that Kalman filters are well suited to the prediction of low frequency univariate financial time series, i.e. recorded either daily (Gultekin and Paisley, 2017) or even down to the resolution of a second. Micro-structural effects start to dominate the signal at higher frequency, typically sub-millisecond level. Knowledge of the market micro-structure becomes fundamental for accurate prediction of price movements and the problem of time series prediction does not fit squarely into the modeling of a time-varying signal.

Modern financial exchanges facilitate the electronic trading of instruments through an instantaneous double auction. At each point in time, the market demand and the supply can be represented by an electronic limit order book, a cross section of orders to execute at various price levels away from the market price. The market price is closely linked to its liquidity - that is the immediacy in which the instrument can be converted into cash. The liquidity of markets are characterized by their depth, the total quantity of quoted buy and sell orders about the market price. The liquidity of the market evolves in response to trading activity (Bloomfield et al., 2005); At any point in time, the amount of liquidity in the market can be characterized by the cross-section of book depths.

Liquid markets are attractive to market participants as they permit the near instantaneous execution of large volume trades at the best available price, with marginal price impact. However, sometimes a large market order, or a succession of smaller markets orders, will consume an entire price level. This is why the market price fluctuates in liquid markets - an effect often referred to by practitioners as a 'price-flip'. Price level consumption is followed by an initial widening of the bid-ask spread which quickly reverts as market makers exploit it, leading to a new mid-price.

There appears to be no consensus, however, on the extent to which limit order books convey predictive information. Early seminal papers studying equities, including Glosten (1994); Seppi (1997), state that the limit orders beyond the insider market contain little information. In contrast, several other studies state that such limit orders are informative(Parlour, 1998; Bloomfield et al., 2005; Cao et al., 2009; Zheng et al., 2013; Kearns and Nevmyvaka, 2013; Cont et al., 2014; D. and Pollak, 2014). In particular, Cao et al. (2009) study the information content of a limit-order book from the Australian Stock Exchange. They found that the book's contribution to price discovery is approximately 22% while the remaining comes from the first level data and transaction prices. They also demonstrate that order imbalances between the demand and supply exhibit a statistically significantly relationship to short-term future returns. There is growing evidence that the study of microstructure is critical to finding longer term relations and even cross-market effects (Dobrislav and Schaumburg, 2016).

Many quantities, such as the probability of price movements given the state of the limit order, are relevant for trading and intraday risk management. The complex relation between order book dynamics and price movements has been the focus of econometric and stochastic modeling (see Engle and Russell (1998); Cont et al. (2010, 2014); Cont and de Larrard (2013); Chavez-Casillas and Figueroa-Lopez (2017) and references therein). For analytic tractability, these models assume a data generation process and typically estimate quantities based on asymptotic limits of diffusion processes.

In a Markovian setting, and under further modeling assumptions, such as the treatment of the arrival rate of market orders as a poisson process, homogenous order sizes, and the assumption of independence of cancellations and orders, a probability of an up-tick is derived. However, these modeling assumptions made are likely too strong for describing micro-scale book dynamics (sub 1ms). At this scale, price is not Markovian, increments are neither independent nor stationary and depend on the state of the order book. Attempts to relax the Markovian assumption, using for example the 'heavy traffic' approximation approach (Cont and de Larrard, 2010; Chavez-Casillas and Figueroa-Lopez, 2017) are best suited for meso-scale analysis of the price movements, but not micro-scale.

Guided by the reduced order book models of Cont and de Larrard (2013), our approach selects similar exogenous variables. In particular, we treat queue sizes at each price level as the independent variables. We additionally include properties of market orders, albeit in a form which we have observed to be most relevant to prediction the direction of price movements. In sharp contrast to stochastic modeling, we do not impose conditional distributional assumptions on the independent variables (a.k.a. features) nor assume that price movements are Markovian.

Most of the aforementioned predictive modeling studies rely on regression for explanatory power of liquidity on the *continuous* volume weighted average price (VWAP), a.k.a. 'smart price'. The utility of the smart price is limited for high frequency trading. So called 'market makers' quote limit orders on both sides of the market in attempt to capture the spread. Their inability to pre-empt a price flip, by adjusting their quotes, typically results in adverse price selection and a loss of profit. A change in the smart price does not imply a price-flip. For example, a change in the volume of the best bid quantity will result in a change in the smart price, but not necessarily a change in the mid-price, the latter effect is attributed to price level consumption from incoming market orders. The successful prediction of a price flip, and not the change in smart price, can therefore be directly used to avoid adverse price selection. Sudden price flips in the the tick data are hard to capture with traditional modeling techniques which rely on instantaneous inside market liquidity imbalance alone.

Breiman (Breiman, 2001) describes the two cultures of statistical modeling when deriving conclusions from data. One assumes a data generating process, the latter uses algorithmic models, treating the data mechanism as unknown. Machine learning falls into the algorithmic class of reduced model estimation procedures. It is designed to provide predictors in complex settings where relations between input and output variables are nonlinear and input space is often high dimensional. A number of researchers have applied machine learning methods to the study of limit order book dynamics (Kearns and Nevmyvaka, 2013; Kercheval and Zhang, 2015; Sirignano, 2016; Dixon et al., 2017).

This paper takes an algorithmic approach to predicting the next event price-flip from a short sequence of observations of limit order book depths and market orders. We choose a spatio-temporal representation (Sirignano, 2016; Dixon et al., 2017) of the limit order book combined with history of the market orders as the predictors. Our approach solves a sequence classification problem in which a short sequence of observations of book depths and market orders can be classified into directional mid-price movement. A sequence classifier offers potential significant benefit to market participants. For example, a market maker can use the classifier to continuously adjust the quotes, potentially reducing the likelihood of adverse price selection. Sequence classification has been considered elsewhere in the literature for lower frequency price movement prediction from historical prices (Leung et al., 2000; Dixon et al., 2016). The novelty of our approach therefore arises from the application of a recurrent neural network classifier to a spatio-temporal representation of the limit order book combined with market order history in order to predict price-flips.

The main contribution of this paper is to describe and demonstrate the potential of recurrent neural networks for classifying short-term price movements from limit order books of financial futures. Training a recurrent neural network architecture can be performed with stochastic gradient descent (SGD) which learns the weights and offsets in an architecture between the layers. Drop-out (DO) performs variable selection (Srivastava et al., 2014). RNNs rely on a moderate amount of training time series data together with a flexible architecture to 'match' in and out of sample performance as measured by mean error, area under the curve (AUC) or the F1 score, which is the harmonic mean of precision and recall. Our performance results demonstrate the ability of the RNN to capture the non-linear relationship between the near-term price-flips and a spatio-temporal representation of the limit order book.

The remainder of the paper is outlined as follows. Section 2 provides an introduction to sequence learning and RNN architectures. Section 2.3 describes the training, validating and testing process required to construct a RNN. Section 3 describes the high frequency classification problem and presents a spatio-temporal formulation of the limit order book variables extracted from a market data feed. Section 4 describes the implementation and quantifies the empirical gains using a RNN to capture discontinuities in the mid-price, versus a Kalman Filter method adapted for classification and a logistic regression using S&P500 E-mini futures level II data. The RNN is shown to compare favorably with other classifiers on a spatio-temporal feature set. We further assess the robustness and stability of the RNN over a entire calendar month, where on any given day, the RNN may make millions of predictions. In particular, we characterize the effect of daily retraining the RNN and it's sensitivity to latency between the trade platform and the exchange. Section 5 concludes.

# 2 Machine Learning

Machine learning addresses a fundamental prediction problem: Construct a nonlinear predictor, $\hat{Y}(X)$, of an output, $Y$, given a high dimensional input matrix $X = (X^{(1)}, \ldots, X^{(P)})$ of $P$ variables. Machine learning can be simply viewed as the study and construction of an input-output map of the form

$$Y = F(X) \text{ where } X = (X^{(1)}, \ldots, X^{(P)}).$$

The output variable, $Y$, can be continuous, discrete or mixed. For example, in a classification problem, $F: X \to Y$ where $Y \in \{1, \ldots, K\}$ and $K$ is the number of categories. When $Y$ is a continuous vector and $f$ is a semi-affine function, then we recover the linear model

$$Y = AX + b.$$

## 2.1 Sequence Learning

If the input-output pairs $\mathcal{D} = \{X_t, Y_t\}_{t=1}^N$ are auto-correlated observations of $X$ and $Y$ at times $t = 1, \ldots, N$, then the fundamental prediction problem can be expressed as a sequence prediction problem: construct a nonlinear times series predictor, $\hat{Y}(\mathcal{X})$, of an output, $Y$, using a high dimensional input matrix of $T$ length sub-sequences $\mathcal{X}$:

$$y = F(\mathcal{X}) \text{ where } \mathcal{X}_t = seq_T(X_t) = (X_{t-T+1}, \ldots, X_t)$$

where $X_{t-j}$ is a $j^{th}$ lagged observation of $X_t$, $X_{t-j} = L^j[X_j]$, for $j = 0, \ldots, T-1$. Sequence learning, then, is just a composition of a non-linear map and a vectorization of the lagged input variables. If the data is i.i.d., then no sequence is needed (i.e. $T = 1$), and we recover the standard prediction problem.

## 2.2 Recurrent Neural Networks (RNNs)

RNNs are sequence learners which have achieved much success in applications such as natural language understanding, language generation, video processing, and many other tasks Graves (2013). We will concentrate on simple RNN models for brevity of notation.

A simple RNN is formed by a repeated application of a function $F_h$ to the input sequence $\mathcal{X}_t = (X_1, \ldots, X_T)$. For each time step $t = 1, \ldots, T$, the function generates a hidden state $h_t$ from the current input $X_t$ and the previous output $h_{t-1}$:

$$h_t = F_h(X_t, h_{t_1}) = \sigma(W_h X_t + U_h h_{t_1} + b_h), \tag{1}$$

for some non-linear activation function $\sigma(x)$. As illustrated in Figure 1, this simple RNN is an unfolding of a single hidden layer neural network (a.k.a. Elman network (Elman, 1991)) over all time steps.

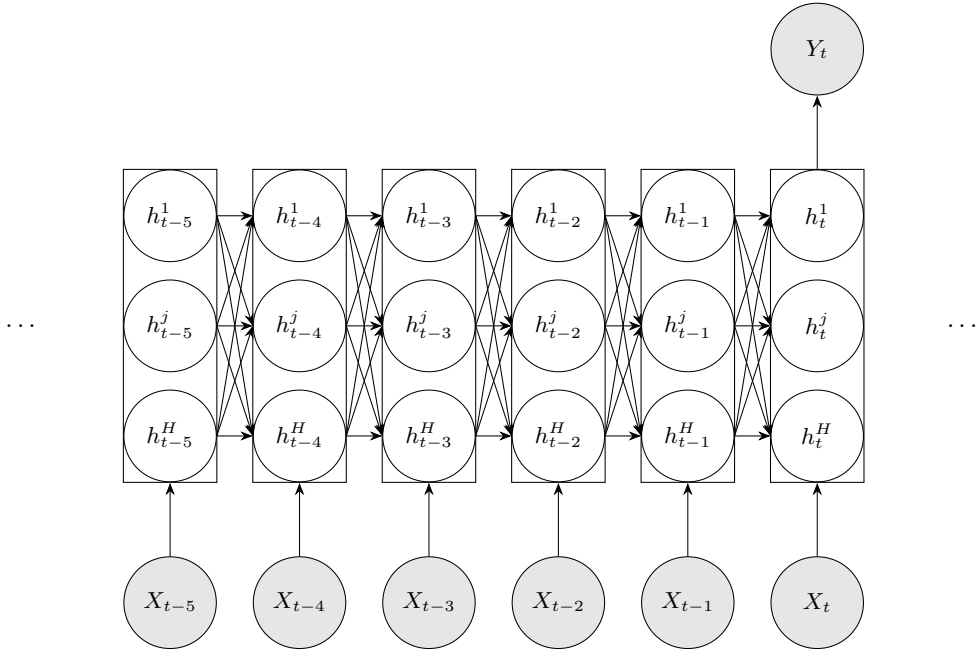

Figure 1: *An illustrative example of a recurrent neural network with one hidden layer, unfolded over a sequence of six time steps. Each input $X_t$ is in the sequence $\mathcal{X}_t$. The hidden layer contains $H$ units and the $j^{th}$ output at time step $t$ is denoted by $h_t^j$. The connections between the hidden units are recurrent and are weighted by the matrix $W_h$. At the last time step $t$, the hidden units connect to a $K$ unit output layer with a 1-of-K output vector $Y_t$.*

When the output is continuous, the model output from the final hidden state, $Y = F_y(h_T)$, is given by the semi-affine function:

$$Y = F_y(h_T) = W_y h_T + b_y, \qquad (2)$$

and when the output is categorical, the output is given by

$$Y = F_y(h_T) = \text{softmax}(F_y(h_T)), \qquad (3)$$

where Y has a 'one-hot' encoding - a K-vector of zeros with 1 at a single position. Here $W = (W_h, U_h, W_y)$ and $b = (b_h, b_y)$ are weight matrices and offsets respectively. $W_h \in \mathbb{R}^{H \times P}$ denotes the weights of non-recurrent connections between the input $X_t$ and the $H$ hidden units. The weights of the recurrence connections between the hidden units is denoted by the recurrent weight matrix $U_h \in \mathbb{R}^{H \times H}$. Without such a matrix, the architecture is simply an unfolded single layer feed-forward network without memory and each observation $X_t$ is treated as an independent observation.

$W_y$ denotes the weights tied to the output of the hidden units at the last time step, $h_t$, and the output layer. If the output variable is a continuous vector, $Y \in \mathbb{R}^M$ then $W_y \in \mathbb{R}^{M \times H}$. If the output is categorical, with K states, then $W_y \in \mathbb{R}^{K \times H}$.

There are a number of issues in the RNN design. How many times should the network being unfolded? How many hidden neurons $H$ in each hidden layer? How to perform 'variable selection'? Many of these problems can be solved by a stochastic search technique, called dropout (Srivastava et al., 2014), which we discuss in Section 2.4.

## 2.3 Training, Validation and Testing

To construct and evaluate a learning machine, we start by controlled splitting of the data into training, validation and test sets. The training data consists of input-output pairs $\mathcal{D} = \{Y_t, X_t\}_{t=1-(T-1)}^{N}$. We then sequence the data to give $\mathcal{D}_{\text{seq}} = \{Y_t, \mathcal{X}_t\}_{t=1}^{N}$.

The goal is to find the machine sequence learner $Y = F(\mathcal{X})$, where we have a loss function $\mathcal{L}(Y, \hat{Y})$ for a predictor, $\hat{Y}$, of the output signal, $Y$. In many cases, there's an underlying probability model, $p(Y \mid \hat{Y})$, then the loss function is the negative log probability $\mathcal{L}(Y, \hat{Y}) = -\log p(Y \mid \hat{Y})$. For example, under a Gaussian model $\mathcal{L}(Y, \hat{Y}) = ||Y - \hat{Y}||^2$ is a $L^2$ norm, for binary classification, $\mathcal{L}(Y, \hat{Y}) = -Y \log \hat{Y}$ is the negative cross-entropy.

In its simplest form, we then solve an optimization problem

$$\underset{W,b}{\text{minimize}}\ f(W,b) + \lambda \phi(W,b)$$

$$f(W,b) = \frac{1}{N} \sum_{t=1}^{N} \mathcal{L}(Y_t, \hat{Y}(\mathcal{X}_t))$$

with a regularization penalty, $\phi(W, b)$.

Here $\lambda$ is a global regularization parameter which we tune using the out-of-sample predictive mean-squared error (MSE) of the model on the verification data. The regularization penalty, $\phi(W, b)$, introduces a bias-variance tradeoff. $\nabla \mathcal{L}$ is given in closed form by a chain rule and, through back-propagation on the unfolded network, the weight matrices $\hat{W}$ are fitted with stochastic gradient descent. See Rojas (1996); Graves (2013) for a further description of stochastic gradient descent as it pertains to recurrent neural networks.

A significant factor raising the appeal of recurrent neural networks to practitioners is `TensorFlow` Abadi et al. (2016), an interface for easily expressing machine learning, and in particular deep learning, algorithms and mapping compute intensive operations onto a wide variety of different hardware platforms and in particular GPU cards. Recently, `TensorFlow` has been augmented by `Edward` Tran et al. (2017) to combine concepts in Bayesian statistics and probabilistic programming with deep learning.

## 2.4 Predictor Selection and Dropout

Dropout is a model or variable selection technique. The input space $\mathcal{X}$, needs dimension reduction techniques which are designed to avoid over-fitting in the training process. Dropout works by removing input variables in $X_t$ randomly with a given probability $\theta$. The probability, $\theta$, can be viewed as a further hyper-parameter (like $\lambda$) which can be tuned via cross-validation. Heuristically, if there are $P = 100$ variables in $X_t$, then a choice of $\theta = 0.1$ will result in a search for models with 10 variables. The dropout architecture with stochastic search for the predictors can be used

$$D_i \sim \text{Ber}(\theta),$$
$$\tilde{X}_t = D \star X_t,\ t = 1, \ldots, T,$$
$$h_t = F_h \left( W_h \tilde{X}_t + U_h h_{t-1} + b_h \right).$$

Effectively, this replaces the input $X_t$ by $D \star X_t$, where $\star$ denotes the element-wise product and $D$ is a 'drop-out operator' - a vector of independent Bernoulli, Ber($\theta$), distributed random variables. The overall objective function is closely related to ridge regression with a g-prior (Heaton et al., 2017). Note that drop-out is not applied to the recurrent connections, only the non-recurrent connections. Graves (2013) provides evidence of the success in RNNs by applying drop-out only to the non-recurrent connections in a LSTM.

# 3 High Frequency Trading

To further motivate why classification is useful for high frequency trading, we shall consider the following instructive scenario. A market maker offers to sell at price $x_t + s/2$ and buy at price $x_t - s/2$ in an attempt to capture the spread $s$, where $x_t$ is the mid-price at time $t$. The market maker runs the risk of an adverse mid-price movement between the fill of the sell order and the buy order. For example, a buy market order matching with the entire resting ask quantity, at price $x_t + s/2$, effectively results in an up-tick of the mid-price, $x_{t+1} = x_t + s$. The market maker has sold at $x_t + s/2$ but can very likely not buy back at $x_t - s/2$ as the bid is no longer at the inside market. Instead the market marker may systematically be forced to cancel the bid and buy back at a price higher than $x_t - s/2$, taking a loss.

Figure 2 (left) illustrates a typical mechanism resulting in an adverse price movement. A snapshot of the limit order book at time $t$, before the arrival of a market order, and after at time $t+1$ are shown in the left and right panels respectively. The resting orders placed by the market marker are denoted with the '+' symbol- red denotes a buy limit order and blue denotes a sell limit order. A buy market order subsequently arrives and matches the entire resting quantity of best ask quotes. Then at event time $t+1$ the limit order book is updated - the market maker's sell limit order has been filled (blue minus symbol) and the buy order now rests away from the inside market. The mechanism is analogous for a down-tick of the book.

In the above illustrated example, the market maker is assumed to have no knowledge of the future and therefore does not pre-empt the price movement by adjusting their quotes accordingly. But suppose now that the market maker reasonably suspects an up-tick and has the capacity to avoid a loss resulting from the adverse price movement.

Figure 2 (right) illustrates the result of a market maker acting to prevent losing the spread -the sell limit order is adjusted to a higher ask price. In this illustration, the buy limit order is not replaced and the market maker may capture a tick more than the spread. However, it's unlikely that the buy limit order will be filled and hence may also need to be adjusted up to the inside market, albeit with loss of queue position. Such trade-offs are the focus of execution optimization and, although important, are not considered further here.

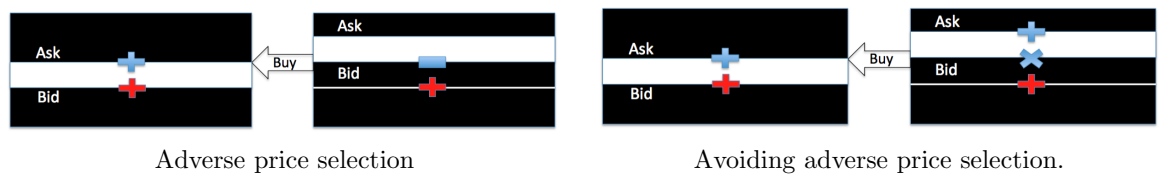

Adverse price selection | Avoiding adverse price selection.

Figure 2: *(left) Illustration of a typical mechanism resulting in adverse price selection. A snapshot of the limit order book is taken at time $t$. Limit orders placed by the market marker are denoted with the '+' symbol- red denotes a buy limit order and blue denotes a sell limit order. A buy market order subsequently arrives and matches the entire resting quantity of best ask quotes. Then at event time $t+1$ the limit order book is updated. The market maker's sell limit order has been filled (blue minus symbol) and the buy order now rests away from the inside market. (Right) A pre-emptive strategy for avoiding adverse price selection is illustrated. The sell limit order is requoted at a higher ask price. In this case, the buy limit order is not replaced and the market maker may capture a tick more than the spread if both orders are filled.*

## 3.1 High Frequency Data

Our dataset is an archived Chicago Mercantile Exchange (CME) FIX format message feed captured from August 1, 2016 to August 31, 2016. This message feed records all transactions in the front month E-mini S&P 500 (ESU6) between the times of 12:00pm and 22:00 UTC. We extract details of each limit order book update, including the nano-second resolution time-stamp, the quoted price and depth for each limit order book level.

The E-mini S&P 500 (ES) tick size is a quarter of a point, or 12.50 per contract. Figure 3 illustrates the intuition behind a typical mechanism resulting in mid-price movement. We restrict consideration to the top five levels of the ES futures limit order book, even though there are updates provided for ten levels. The chart on the left represents the state of the limit order book prior to the arrival of a sell market order (a.k.a. 'aggressor'). The x-axis represents the price levels and the y-axis represents the depth of book at each price level. Red denotes bid orders and blue denotes ask orders. The highest bid price ('best bid') is quoted at $2175.75 with a depth of 103 contracts. The second highest bid is quoted at $2175.5 with a depth of 177 contracts. The lowest ask ('best ask' or 'best offer') is quoted at $2176 with 82 contracts and the second lowest ask is quoted at $2176.25 with 162 contracts.

The chart on the right shows the book update after a market order to sell 103 contracts at $2175.75. The aggressor is sufficiently large to match all of the best bids. Once matched, the limit order is updated with a lower best bid of $2175.5. The gap between the best ask and best bid would widen if it weren't for the arrival of 23 new contracts offered at a lower ask price of $2175.75. The net effect is a full down-tick of the mid-price.

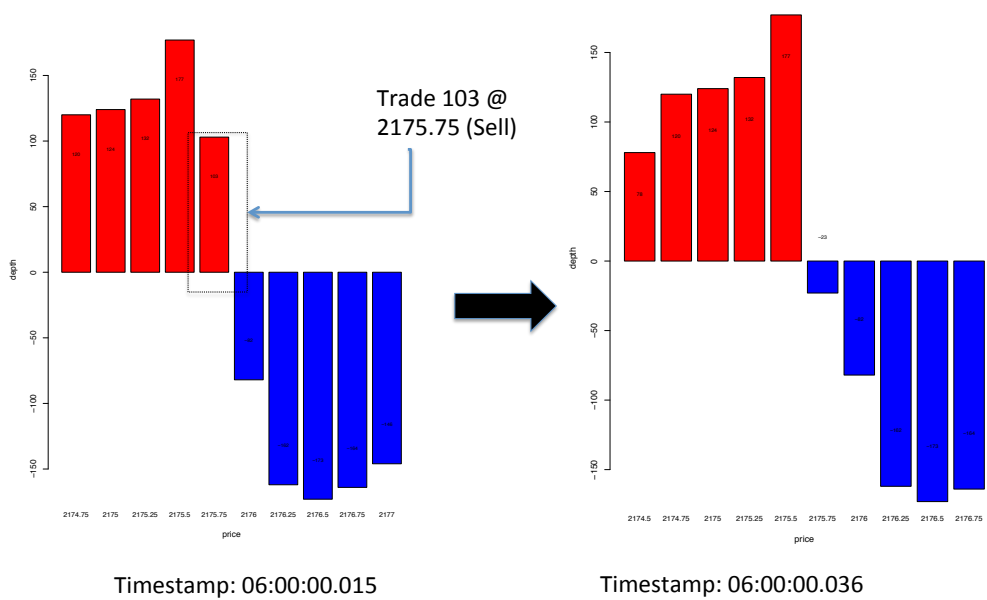

Figure 3: *This figure illustrates a typical mechanism resulting in mid-price movement. The charts on the left and right respectively show the limit order book before and after the arrival of a large sell aggressor. The aggressor is sufficiently large to match all of the best bids. Once matched, the limit order is updated with a lower best bid of $2175.5. The gap between the best ask and best bid would widen if it weren't for the arrival of 23 new contracts offered at a lower ask price of $2175.75. The net effect is a full down-tick of the mid-price.*

Table 1 shows the limit order book before and after the arrival of the sell aggressor. The response is mid-price movement, in units of ticks, over the subsequent interval. $p_{i,t}^b$ and $d_{i,t}^b$ denote the level $i$ quoted bid price and depth of the limit order book at time $t$. $p_{i,t}^a$ and $d_{i,t}^a$ denote the corresponding level $i$ quoted ask price and depth. Level $i = 1$ corresponds to the best ask and bid prices. The mid-price at time $t$ is denoted by

$$p_t = \frac{p_{1,t}^a + p_{1,t}^b}{2}. \tag{4}$$

This mid-price can evolve in minimum increments of half a tick but almost always is observed to move at increments of a tick over time intervals of a milli-second or less.

| Timestamp | $p_{1,t}^b$ | $p_{2,t}^b$ | ... | $d_{1,t}^b$ | $d_{2,t}^b$ | ... | $p_{1,t}^a$ | $p_{2,t}^a$ | ... | $d_{1,t}^a$ | $d_{2,t}^a$ | ... | Response |
|---|---|---|---|---|---|---|---|---|---|---|---|---|---|
| 06:00:00.015 | 2175.75 | 2175.5 | ... | 103 | 177 | ... | 2176 | 2176.25 | ... | 82 | 162 | ... | -1 |
| 06:00:00.036 | 2175.5 | 2175.25 | ... | 177 | 132 | ... | 2175.75 | 2176 | ... | 23 | 82 | ... | 0 |

Table 1: *This table shows the limit order book of ESU6 before and after the arrival of the sell aggressor listed in Figure 3. The response is the mid-price movement over the subsequent interval, in units of ticks. $p_{i,t}^b$ and $d_{i,t}^b$ denote the level $i$ quoted bid price and depth of the limit order book at time $t$. $p_{i,t}^a$ and $d_{i,t}^a$ denote the corresponding level $i$ quoted ask price and depth.*

Using a model to predict the next mid-price movement is well suited to futures markets which often experience a consistent surge of activity at particular times when participants enter into and close out their futures' positions. The importance of predicting in event time can be observed from the hourly limit order book rates shown in Figure 4. In particular, a surge of quote adjustment and trading activity is consistently observed between the hours of 7-8am CST and 3-4pm CST. At these times, book updates may occur as often as every micro-second, where as at other times of the day, the book updates may occur every tenth of a milli-second. Put differently, the 1ms ahead prediction of the mid-price may be relevant for avoiding adverse selection at off-peak periods of the day, and less relevant at peak periods.

Each limit order book update is recorded as an observation. The result of categorizing (a.k.a. labeling) each observation leads to a class imbalance problem, as approximately 99.9% of the observations have a zero response. To construct a 'balanced' training set, observations (sequences of input variables) labeled by the minority class are oversampled with replacement and the majority class observations are undersampled without replacement.

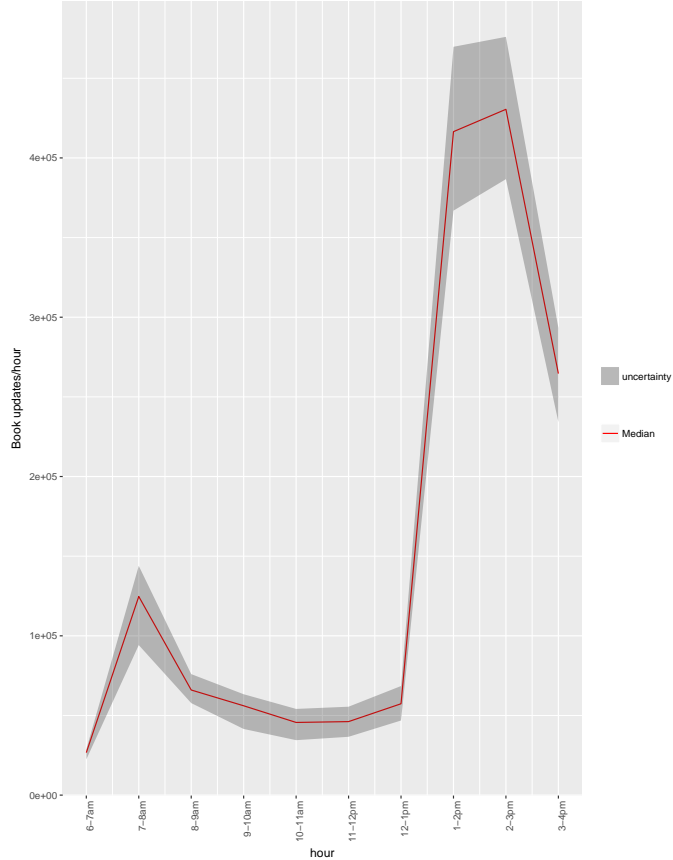

Figure 4: *The hourly limit order book rates of ESU6 are shown by time of day. A surge of quote adjustment and trading activity is consistently observed between the hours of 7-8am CST and 3-4pm CST.*

Partially following Kercheval and Zhang (2015), we compose our feature set of five levels of prices, volumes and number of limit orders on both the ask and bid side of the book. We additionally, and somewhat heuristically via a process of 'feature engineering', characterize order flow by the ratio of the number of market buy orders arriving in the prior 50 observations to the resting number of ask limit orders at the top of book. We construct the analogous ratio for the sell by orders. This rationale for this ratio is motivated by our observation that an increase in this ratio will more likely deplete the best ask level and the mid-price will up-tick, and vice-versa for a down-tick. The combination of this spatial representation of the limit order book and the order flow gives a total of $P = 32$ features.

## 4 Results

The exact architecture and weight matrix sizes of our recurrent neural network are given by

$$\text{output}: Y^k = \text{softmax}(F_y^k(h_T)) = \frac{\exp(F_y^k(h_T))}{\sum_{j=1}^{K} \exp(F_y^j(h_T))},$$

$$\text{hidden states}: h_t = \max\left(W_h X_t + U_h h_{t-1} + b_h, 0\right), \quad t = 1, \dots, T,$$

|         | $\hat{Y} = -1$ | | | $\hat{Y} = 0$ | | | $\hat{Y} = 1$ | | | |
|---------|-----------|--------|-------|-----------|--------|-------|-----------|--------|-------|-----------|
|         | precision | recall | f1    | precision | recall | f1    | precision | recall | f1    | size      |
| mean    | 0.092     | 0.830  | 0.165 | 0.996     | 0.793  | 0.881 | 0.090     | 0.838  | 0.161 | 99504.720 |
| std.dev | 0.031     | 0.077  | 0.050 | 0.004     | 0.075  | 0.050 | 0.027     | 0.080  | 0.045 | 68034.218 |

Table 2: *This table summarizes the performance of the RNN over a 20 trading day test period. The RNN classifier is used to predict the categories: $\{Y = -1, Y = 0, Y = 1\}$. The mean and standard deviation of the daily precision, recall and f1 scores of each model are shown together with the average and standard deviation of the training set sizes.*

where $W_h \in \mathbb{R}^{20 \times 32}, U_h \in \mathbb{R}^{20 \times 20}$ and $W_y \in \mathbb{R}^{3 \times 20}$. We initialize the hidden states to zero. We use the SGD method, implemented in `Python`'s `TensorFlow` Abadi et al. (2016) framework, to find the optimal network weights, bias terms and regularization parameters. We employ an exponentially decaying learning rate schedule with an initial value of $10^{-2}$. The optimal $\ell_2$ regularization is found, via a grid-search, to be $\lambda_2 = 0.01$. The Glorot and Bengio method is used to initialize the weights of the network Glorot and Bengio (2010).

Times series cross-validation is performed over 20 consecutive trading days. Training sets are compiled from the previous 3 trading days and contain, on average, $5,192,822$ observations. These sets are balanced resulting in a reduced training set size of typically just less than $100,000$ observations. The validation and test sets are compiled for the next trading day following the 3 day training period. These are unbalanced, with the verification set containing $2 \times 10^5$ observations and the remaining test set, on average, containing approximately $1.5 \times 10^6$ observations. Each experiment is run for 1000 epochs with a mini-batch size of 500 drawn from the training set of 32 input variables. We follow the standard convention of choosing the number of epochs based on convergence of the cross-entropy and the mini-batch size is chosen for computational performance. Each sequence is chosen to be of length 10. The gridded search to find the optimal network architecture and regularization parameters takes several hours on a modern graphics processing unit (GPU). The search yields several candidate architectures and parameter values.

Table 2 summarizes the performance of the RNN over 20 trading days. To reliably measure performance on the unbalanced test set, we compute the $F1$ score - the geometric mean of the precision and recall:

$$F1 = 2 \frac{\text{precision} \cdot \text{recall}}{\text{precision} + \text{recall}}.$$

The $F1$ score is designed for binary classification problems. When the data has more than two classes, the $F1$ score is provided for each class $k$ by setting $Y = k$ to the positive and all remain classes to the negative label. The score is highest for the zero label corresponding to a prediction of a stationary mid-price over the next interval. The $F1$ scores for a predicted up-tick $F1(1)$ and down-tick $F1(-1)$ are also shown. We observe that the performance of the RNN when $Y = -1$ and when $Y = 1$ are the positive labels is comparable. As expected, the RNN can predict these classes with comparable recall but drastically less precision than for the $Y = 0$ positive label model. The average and standard deviation of the size of balanced training sets are shown by the `size` column.

Table 3 studies the effect of the number of time steps in the RNN model by performing time series cross-validation over the 20 training days. The performance is observed to increase as the number of time steps is increased from 1 to 10. This provides evidence for the need for recurrence in the neural network. While the difference between 10 and 20 steps is marginal, a sharp decline in performance is observed at 50 and 100 time steps. This suggests that short-term memory is adequate for sequence classification. Although not shown here, we also observe that the number of hidden units should be between 10 and 20.

Figure 5 shows some of the predictor variables used for the model comparison in Table 4. The black line represents the observed change in mid-price over a 34 milli-second period from 16:37:52.560 to 16:37:52.594 on August 4th, 2016. We observe that the mid-price typically moves in two successive half-ticks. A price level is first consumed and, with some small delay, a new price level is subsequently quoted. The liquidity imbalance (blue), scaled here to the $[-1, 1]$ interval, although useful in predicting the direction of the next occurring price change, is generally a poor choice for predicting when the price change will occur. It is clear from the results in Table 4 and the visualization here that it's inadequate for next-event price prediction. This order flow approximation is a better predictor of next-event price movement, although used alone is difficult to interpret when either of the buy (red) and sell order flows (green) are small or comparable in size. Hence, in the spatio-temporal feature set, we combine the order flow approximations with the liquidity imbalance, mid-prices and depths at other price levels.

| n_steps | $\hat{Y} = -1$ | $\hat{Y} = 0$ | $\hat{Y} = 1$ |
|---|---|---|---|
| 1 | 0.089 | 0.528 | 0.091 |
| 2 | 0.097 | 0.837 | 0.103 |
| 5 | 0.131 | 0.848 | 0.120 |
| 10 | 0.165 | 0.881 | 0.161 |
| 20 | 0.163 | 0.878 | 0.162 |
| 50 | 0.151 | 0.836 | 0.144 |
| 100 | 0.118 | 0.807 | 0.115 |

Table 3: *This table shows the effect of the number of time steps in the RNN model on the average F1 score, as measure by time series cross-validation over the 20 testing days.*

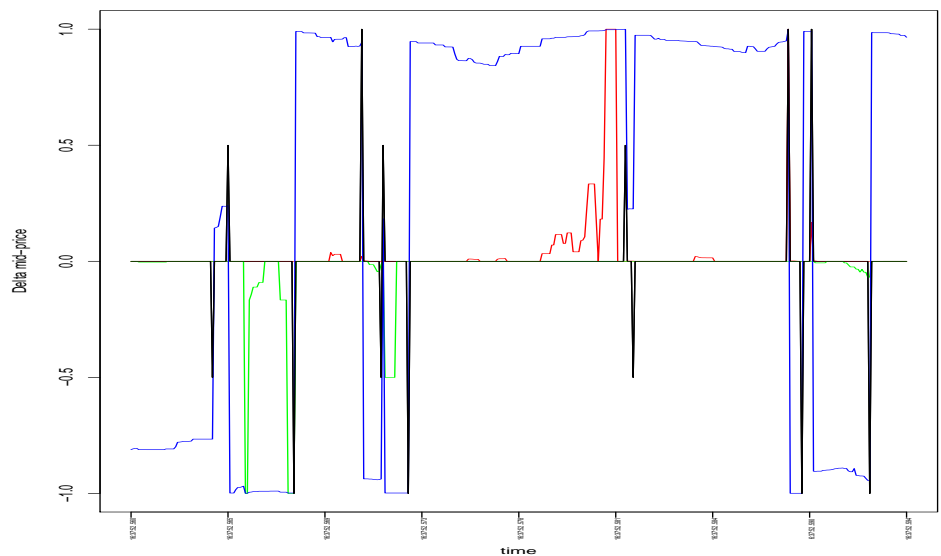

Figure 5: *This figure shows some of the predictor variables used for the model comparison in Table 4. The black line represents the observed change in mid-price over a 34 milli-second period from 16:37:52.560 to 16:37:52.594. The liquidity imbalance (blue), scaled here to the $[-1, 1]$ interval, although useful in predicting the direction of the next occurring price change, is generally a poor choice for predicting when the price change will occur. This order flow is a better predictor of next-event price movement, although is difficult to interpret when either of the buy (red) and sell order flows (green) are small.*

Table 4 compares the performance of a various classifiers on the following feature sets: (i) instantaneous liquidity imbalance using the best bid to ask ratio; (ii) instantaneous order flow using the liquidity imbalance, volume of buy or sell aggressors to resting quantity and mid-price; and (iii) the spatio-temporal representation of the order book combined with the history of the aggressors. We use uniform white noise as a control for the performance comparisons. First, the RNN is compared with a logistic regression (LR) and a linear Kalman filter (KF) model designed for observations of $Y$ from the exponential family of distributions, for which the Bernoulli distribution is a member. We emphasis that no history is incorporated in either model applied to the first two feature sets. As the latter two models only support the modeling of Bernoulli random variables, we follow a standard many-model representation of a multi-classifier, training 3 separate binary classifiers to represent a 3-state output. The probability threshold for predicting each positive state is set to 0.5.

The LR model is implemented in the `glm` R package and the KF model is implemented in the `KFAS` R package (Helske, 2017). The KF model approximates the posterior density by simulating 1000 trials over a section of a single test set of size $108,035$ and is considerably more compute and memory intensive to evaluate than the other classifiers. The recall and F1 scores of the KF model applied to the liquidity imbalance predictor, and separately to the order flow predictors, are all higher than the RNN and LR models, although some of the precisions are not. All methods generally perform considerably better with order flow predictors than liquidity imbalance alone.

The spatio-temporal feature set contains 32 predictors, each lagged by up to ten times. To avoid a multi-collinearity problem with including multiple lagged features in a LR model, we use the elastic net method, implemented in the `glmnet` R package (Friedman et al., 2010; Simon et al., 2011). The KF model does not scale to this feature set. The RNN is compared with a two layer feed-forward architecture with 200 hidden units in the first hidden layer and 100 hidden units in the second. Feed-forward architectures are not designed explicitly for time series and so, as with the elastic net method, we ensure consistency in comparison with the RNN by lagging each of the features up to lag 9 resulting in 320 input variables. The choice of tuning parameters for the regularization and learning rate can lead to large performance variations. The results reported here using an initial learning rate of 0.01 and $\lambda = 0.1$, although several candidate values exist. In separate experiments, not reported here, we observe that the performance of a deep feed-forward architecture can perform as well as a RNN, however with so many more hidden units, the approach is more computationally intensive and less robust to the choice of initial parameters.

| Features | Method | $\hat{Y} = -1$ | | | $\hat{Y} = 0$ | | | $\hat{Y} = 1$ | | |
|---|---|---|---|---|---|---|---|---|---|---|
| | | precision | recall | f1 | precision | recall | f1 | precision | recall | f1 |
| Liquidity Imbalance | Logistic | 0.010 | 0.603 | 0.019 | 0.995 | 0.620 | 0.764 | 0.013 | 0.588 | 0.025 |
| | Kalman Filter | 0.051 | 0.540 | 0.093 | 0.998 | 0.682 | 0.810 | 0.055 | 0.557 | 0.100 |
| | RNN | 0.037 | 0.636 | 0.070 | 0.996 | 0.673 | 0.803 | 0.040 | 0.613 | 0.075 |
| Order Flow | Logistic | 0.042 | 0.711 | 0.079 | 0.991 | 0.590 | 0.740 | 0.047 | 0.688 | 0.088 |
| | Kalman Filter | 0.068 | 0.594 | 0.122 | 0.996 | 0.615 | 0.751 | 0.071 | 0.661 | 0.128 |
| | RNN | 0.064 | 0.739 | 0.118 | 0.995 | 0.701 | 0.823 | 0.066 | 0.728 | 0.121 |
| Spatio-temporal | Elastic Net | 0.063 | 0.754 | 0.116 | 0.986 | 0.483 | 0.649 | 0.058 | 0.815 | 0.108 |
| | RNN | 0.084 | 0.788 | 0.153 | 0.999 | 0.729 | 0.843 | 0.075 | 0.818 | 0.137 |
| | FFWD NN | 0.066 | 0.758 | 0.121 | 0.999 | 0.657 | 0.795 | 0.065 | 0.796 | 0.120 |
| | White Noise | 0.004 | 0.333 | 0.007 | 0.993 | 0.333 | 0.499 | 0.003 | 0.333 | 0.007 |

Table 4: *This table shows the performance results of various classifiers on the following feature sets: (i) instantaneous liquidity imbalance using the best bid to ask ratio; (ii) instantaneous order flow using the liquidity imbalance, volume of buy or sell aggressors to resting quantity and mid-price; and (iii) the spatio-temporal representation of the order book combined with the history of the aggressors. We use uniform white noise as a control for the performance comparisons. By setting the positive to $\{Y = -1, Y = 0, Y = 1\}$, the precision, recall and f1 score are each recorded over a single test set of size 108,035.*

Figure 6 compares the RNN and KF model estimated probabilities of a next event up-tick (red) or down-tick (green) over a 34 milli-second period from 16:37:52.560 to 16:37:52.594 on August 4th, 2016. The top graph shows the estimated probabilities from a Kalman filter with just liquidity imbalance as the predictor. The black line represents the observed change in mid-price. The middle graph shows the estimated probabilities using a Kalman filter with the predictors: liquidity imbalance, mid-price and ratio of market orders to resting quantity. The bottom graph shows a RNN using the spatio-temporal order book representation and ratio of market orders to resting quantity. We observe that the RNN often correctly predicts the up-tick movement but also falsely predicts movements when the mid-price is neutral. These false positives (or negatives), when the observed mid-price is stationarity, oftentimes result in unnecessary order cancellations and loss of queue position. While the false positive rates can be reduced by increasing the threshold probability for detecting a price movement, this may result in increased adverse price selection - the false prediction of a stationary mid-price when the observed mid-price flips. Thus either the probability or the probability threshold should be used in a market marking strategy.

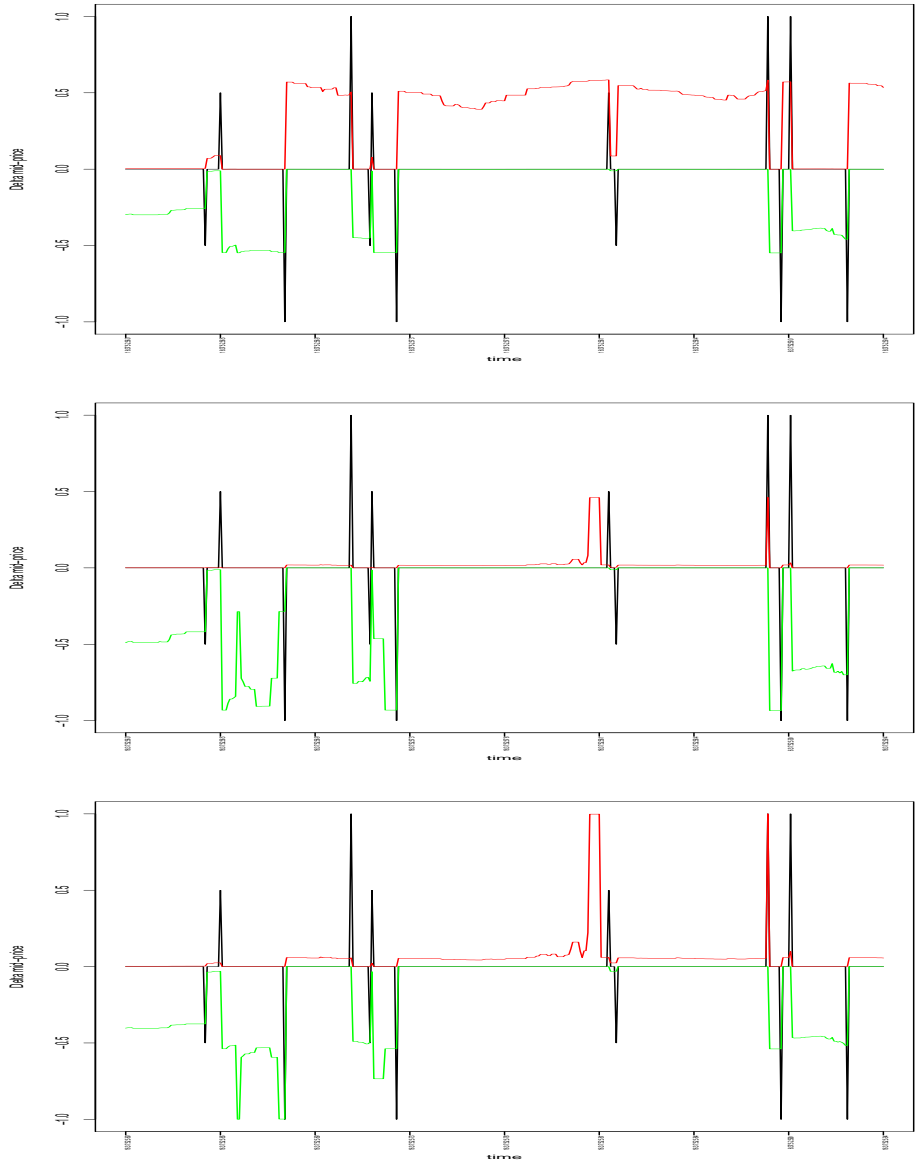

Figure 6: *(top) Estimated probabilities of next-event up-ticks (red) or down-ticks (green) of the book using a Kalman filter with just liquidity imbalance as the predictor. The black line represents the observed change in mid-price over a 34 milli-second period from 16:37:52.560 to 16:37:52.594. (Middle) Estimated probabilities of next event up-ticks (red) or down-ticks (green) using a Kalman filter with the predictors: liquidity imbalance, mid-price and ratio of market orders to resting quantity. (Bottom) Estimated probabilities of next event up-ticks (red) or down-ticks (green) from a RNN using the spatio-temporal order book representation and ratio of market orders to resting quantity.*

Figure 7 shows the F1 scores of the RNN, measured on any hourly basis. We observe little variation in the F1 scores for up-tick and down-tick predictions and hence no evidence of secular F1 score decay and hence the need to retrain the model intra-day. In fact when we observe the opposite from the F1 score for predicting no book movement- the F1 score is observed to gradually increase over the course of the day.

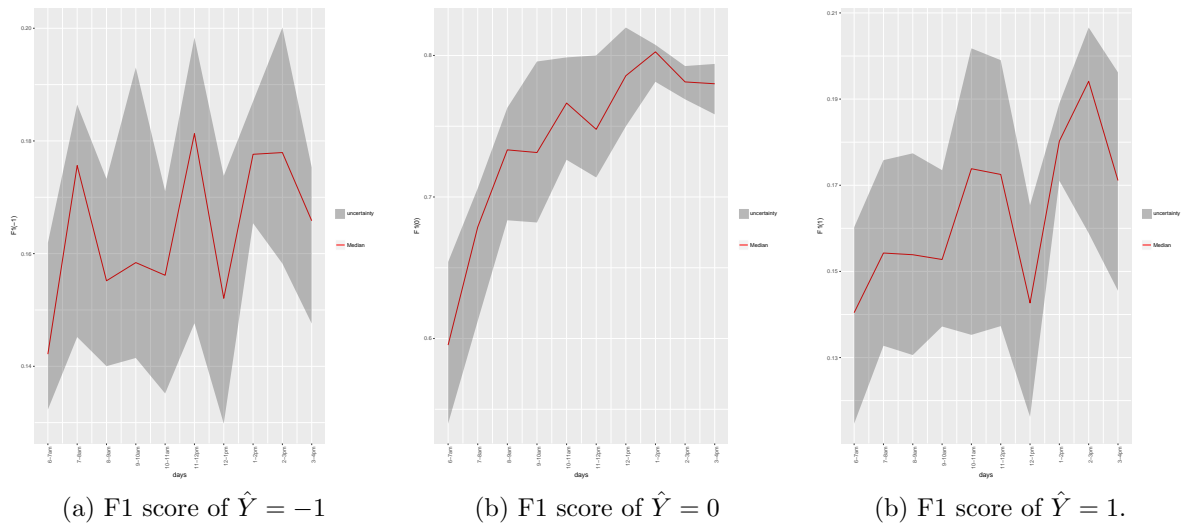

(a) F1 score of $\hat{Y} = -1$    (b) F1 score of $\hat{Y} = 0$    (b) F1 score of $\hat{Y} = 1$.

Figure 7: *The intra-day F1 scores are shown for (left) downward, (middle) neutral, or (right) upward next price movement prediction.*

Figure 8 shows the importance of retraining the model on a daily basis. The F1 scores are observed to exhibit a secular decay when the model is not re-trained on a daily basis but instead trained only at the beginning of the month. While the evidence for secular decay is clear, it is surprising to observe that a model trained at the beginning of the month can still be effective weeks later. Typically econometrics models for non-stationary time series perform very poorly when not frequently re-fitted to new data. This surprising model robustness, without re-training, suggests a robust relationship between the liquidity imbalance and price movements.

Figure 9 compares the Receiver Operator Characteristic (ROC) curves of a binary RNN classifier over varying prediction horizons. The plot is constructed by varying the probability threshold (a.k.a. cut-points) for positive classification $Y = 1$ over the interval $[0.5, 1)$ and estimating the true positive and true negative rate of each model. The dashed line shows the performance of a white-noise classifier. We observe that the performance of the RNN decays as we increase the prediction horizon from the next book update event to 1s in the future. Latency between trading systems and the exchange is dependent on several factors including the execution platform, the distance of the co-located server to the exchange and even the amount of incoming traffic to the CME matching engines. It is commonplace however for this latency to lie in the range of 100 micro-seconds to 1ms. Our estimate, based on preliminary investigation, is that a c/c++ implementation of our RNN can predict in around 100 micro-seconds on a modern CPU with compiler optimizations but without low level programming techniques for code optimization.

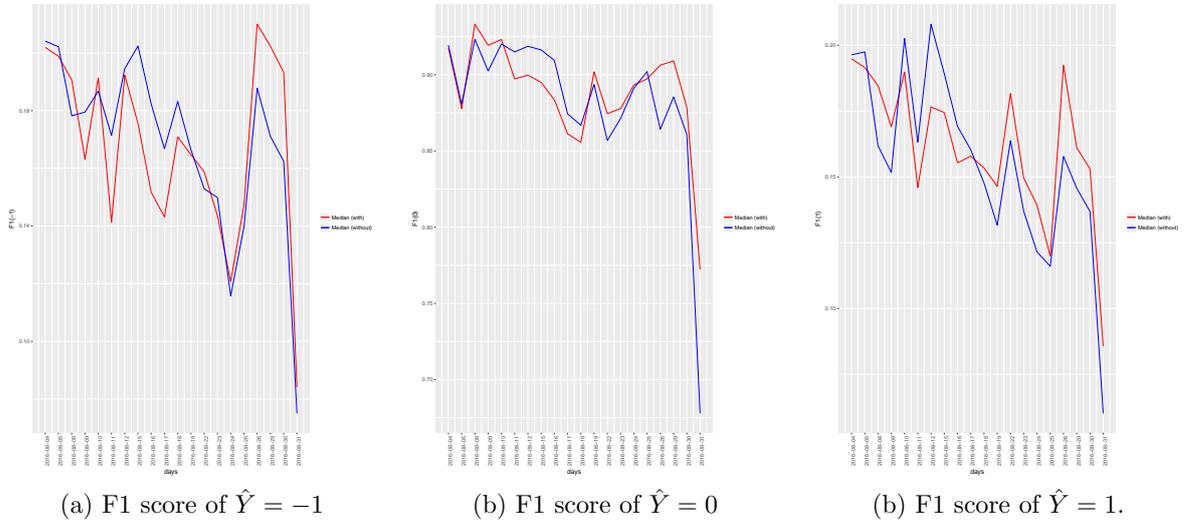

(a) F1 score of $\hat{Y} = -1$  (b) F1 score of $\hat{Y} = 0$  (b) F1 score of $\hat{Y} = 1$.

Figure 8: *The F1 scores over the calendar month, with (red) and without (blue) daily retraining of the RNN, are shown for (left) downward, (middle) neutral, or (right) upward next price movement prediction.*

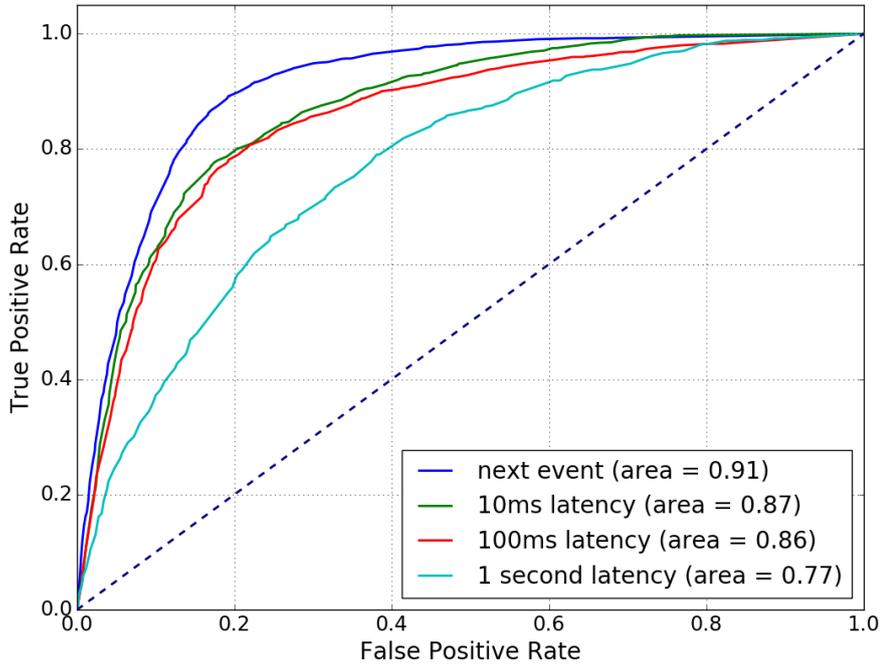

Figure 9: *The Receiver Operator Characteristic (ROC) curves of a binary RNN classifier over varying prediction horizons. In practice, the prediction horizon should be chosen to adequately account for latency between the trade execution platform and the exchange.*

## 5 Conclusion

Recurrent neural networks (RNNs) are types of artificial neural networks (ANNs) that are well suited to forecasting and sequence classification. This paper solves a sequence classification problem in which a short sequence of observations of limit order book depths and market orders is used to predict a next event price-flip. The capability to adjust quotes according to this prediction reduces the likelihood of adverse price selection. Our results demonstrate the ability of the RNN to capture the non-linear relationship between the near-term price-flips and a spatio-temporal representation of the limit order book.

Our assessment, based on CME listed ESU6 level II data over the month of August 2016, is that RNNs are well suited to expressing the non-linear relationship when the next event mid-price movement and very recent history of the liquidity imbalance, order flow and other properties of the limit order book. Since longer memory in our features appears negligible, we speculate that there is little to no benefit in using a LSTM. In order for the RNN to be implemented for use in a HFT market making strategy, we recommend pursing the following: (i) first design a market making strategy that is profitable after accounting for additional latency resulting from the compute time of the RNN prediction; (ii) assess the profit impact of false positives and false negatives during backtesting; and (iii) commensurately extend the strategy to 'weight' the model probabilities of a book flip with loss of queue position resulting from a cancellation.